# Adopting RAG for LLM-Aided Future Vehicle Design

Vahid Zolfaghari*, Nenad Petrovic*, Fengjunjie Pan*, Krzysztof Lebioda*, Alois Knoll*
*Technical University of Munich, Robotics, Artificial Intelligence and Embedded Systems*, Munich, Germany
Email: v.zolfaghari@tum.de, nenad.petrovic@tum.de, f.pan@tum.de, krzysztof.lebioda@tum.de, k@tum.de

*Abstract*— In this paper, we explore the integration of Large Language Models (LLMs) with Retrieval-Augmented Generation (RAG) to enhance automated design and software development in the automotive industry. We present two case studies: a standardization compliance chatbot and a design copilot, both utilizing RAG to provide accurate, context-aware responses. We evaluate four LLMs—GPT-4o, LLAMA3, Mistral, and Mixtral—comparing their answering accuracy and execution time. Our results demonstrate that while GPT-4 offers superior performance, LLAMA3 and Mistral also show promising capabilities for local deployment, addressing data privacy concerns in automotive applications. This study highlights the potential of RAG-augmented LLMs in improving design workflows and compliance in automotive engineering.

## I. Introduction

Large Language Models (LLMs) are currently trending topic in area of Artificial Intelligence (AI) and their adoption in practice is becoming more and more relevant every day. Despite that their question answering power enriches the end-user experience for many applications and services, they also aim to provide enhancements to creative processes, such as design and engineering decisions.

However, the adoption of such LLM-based approach in some engineering domains would assume availability of sensitive information, such as trade secrets, customer's private/personnel data and technical details which should be only available internally to employees and never exposed outside the boundaries of the underlying organizations. Therefore, the real potential of leveraging LLMs for many innovative usage scenarios is reduced due to these reasons.

It is already shown that LLMs can bring many benefits when it comes to software development, such as providing suggestions, analysis source code or even generating executable code [1], [2], [3]. However, in areas such as automotive, both hardware and software requirements can be also covered by non-disclosure policies and other mechanisms making the utilization of the necessary information as input to LLM quite concerning for the companies [2]. Initially, most of the LLM-driven services were based on models including billions of parameters, which requires enormously powerful cloud infrastructure and were not aimed to be run locally. In almost all cases, they were run on large provider's infrastructure, such as Amazon and Google. Therefore, providing documents and other specification-related information in automotive was considered quite problematic - as it involved practically sending the secret information to online services and other parties involved. For that reason, the adoption of LLM-aided tools in automotive was considered quite problematic and has not been the main focus of researchers exploring the usage of LLMs.

In order to overcome the mentioned issue of giving out the secret information to other parties while using LLM-based services, there are two potential solutions. The first one is to entirely deploy the LLM-based service within the boundaries of the organization, relying on local infrastructure (such as high-performance computing clusters). However, this approach would involve additional costs for both the hardware needed (such as high-performance GPUs) and efforts for setting up the clusters and their proper configuration. Moreover, the LLM model itself would need to be trained or fine-tuned using the company's non-disclosed (secret) data, which can require enormous amount of time (depending on the amount of data). Finally, cost-effective local clusters are usually only suitable for smaller-scale LLMs (reduced number of parameters), which might not achieve satisfiable results for the aimed purpose.

On the other side, another solution is to make use of Retrieval Augmented Generation (RAG) [4], [5], which is a technique that aims augmenting LLM knowledge with additional data. Most of the general purpose LLMs can provide quite accurate answers about variety of topics. However, the scope of their knowledge is, in most cases, limited to the publicly available data until the specific point in time when the training was performed. Therefore, in cases when we want to build LLM-enabled applications and services which are able to also consider private data or additional information provided after training, the underlying knowledge of LLM has to be augmented. In that context, the process of retrieval of relevant information and augmenting the prompts targeting LLM is referred to as RAG.

In this work, we aim to adopt RAG-based approach for development of design and software development tools, focusing on bridging the gap of non-disclosable data availability in area of automotive domain. As outcome, we propose vehicular system design workflow making use of LLM and RAG. Furthermore, we present its prototype implementation in context of central car server architecture scenarios. Finally, several widely adopted LLMs will be utilized and their performance compared - taking into account both the outcome quality and additional efforts needed for fine tuning or customization.

## II. Background and Related Works

### A. Retrieval-Augmented Generation (RAG)

Typical RAG-enabled workflow consists of two main processes [4], [5]:

*1) indexing* - involves creation of a pipeline to ingest and index data from a textual source (such as PDF or Word document), typically performed apriori, offline. Furthermore, this phase can be broken down into the following steps: a) *load* - data is initially loaded using document loaders; b) *split* - large documents are broken into smaller chunks with text splitters, making them easier to search and fit within the model's context window; c) *store* - these chunks are stored and indexed, often using a vector store and an embeddings model, for efficient searching later.

*2) retrieval and generation* - runtime process where the system takes a user query, retrieves relevant data from the index, and passes it to a model for generating a response. This phase consists of: a) *retrieve* - relevant data chunks are fetched from storage using a Retriever in response to a user query; b) *generate* - LLM generates an answer by using a prompt that combines the question and the retrieved data.

*B. Langchain and Additional Tools*

For implementation, we rely on Langchain [6] which offers a set of tools making the development of LLM-enabled applications easier.

Concretely, the framework consists of the following open-source libraries: 1) *langchain-core* - offers base abstractions and the *LangChain Expression Language* (LCEL), which is fundamental for many components in LangChain, providing a declarative approach to composing chains. 2) *langchain-community* - Includes third-party integrations such as partner packages like *langchain-openai*; 3) *langchain* - Comprises chains, agents, and retrieval strategies that make up the cognitive architecture of an application; 4) *langgraph* - Supports building robust, stateful multi-actor applications with LLMs by representing steps as edges and nodes in a graph; 5) *langserve* - enables the deployment of LangChain chains as REST APIs

The broader ecosystem includes also *LangSmith*, which is a developer platform for debugging, testing, evaluating, and monitoring LLM applications, seamlessly integrated with LangChain.

*C. Related Works*

In the existing literature, there are several solutions (both scientific and industry-oriented) which adopt LLMs to aid the various aspects of automotive engineering and software development. Table I gives an overview and summarizes the main characteristics of those works.

TABLE I. LLMs AND AUTOMOTIVE: EXISTING WORKS

| Reference | Aspect | Model |
|---|---|---|
| [7] | Hazard Analysis & Risk Assessment (HARA) workflow | ChatGPT |
| [8] | Systems Theoretic Process Analysis (STPA) applied to Automatic Emergency Brake (AEB) and Electricity Demand Side Management (DSM) scenarios | ChatGPT |
| [9] | Chatbot-alike employee advisor and production-related decision making support tool | ChatGPT |
| [7] | Driving assistance based on collected data | ChatGPT |

| Reference | Aspect | Model |
|---|---|---|
| [2] | Code generation based on textual user input, constrained by metamodel | ChatGPT |

As it can be seen, the existing publications mostly present proof-of-concept results based on widely adopted ChatGPT[8]. Despite that it is among the most powerful LLM-based services, there are two main drawbacks which represent huge barrier in area of automotive: 1) API usage and fine tuning are charged on per-token basis, which might cause additional expenses 2) online access is required, which might be a concern for vendors and other relevant parties in the chain of automotive development, as data is sent outside the boundaries of the underlying organizations, which is potentially problematic in many scenarios.

Therefore, in our paper, apart from ChatGPT, we also explore the capabilities of other approved open-source alternative – LLAMA, for its two versions: 2 and 3 (the latest one). Considering that it is less demanding in computing power compared to ChatGPT (but still quite powerful), we can also perform deployment locally, without exposing data. Additionally, most of the approaches aiming the assistance of engineering and analysis in area of automotive, despite that consider taking user feedback into account, mostly do not go further than simple user-based result validation.

III. IMPLEMENTATION OVERVIEW

For the purpose of executing prompts against vendor-specific documents and specifications encapsulating good design practices, we have employed the *Retrieve and Re-Rank* approach as a powerful technique for information retrieval and question-answering systems.

The general process of Retrieve and Re-rank is depicted in Fig. 1.

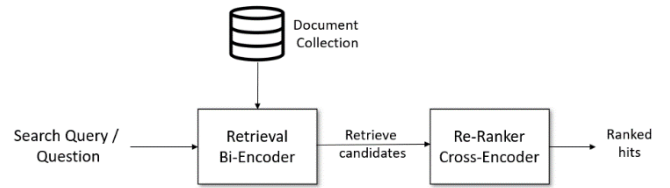

Fig. 1. High-level depiction of Retrieve and Re-rank approach.

This methodology involves two key stages: initial retrieval of a set of potentially relevant documents, followed by a re-ranking process to filter out the most relevant results. The first step in our *Retrieve and Re-Rank* pipeline involves semantic search using a *Bi-Encoder* model. For this purpose, we'll use the *multi-qa-MiniLM-L6-cos-v1*[9] model from *SentenceTransformers*. This sentence-transformers model converts sentences and paragraphs into a 384-dimensional dense vector space, optimized for semantic search. It has been trained on 215 million question-answer pairs from various sources. For the cross-encoder we have used the *cross-encoder/ms-marco-MiniLM-L-6-v2*[10].

We employ a two-stage search process to efficiently identify relevant context for our prompts. Since we're dealing with a large number of text chunks, we leverage the speed of a bi-encoder model in the first stage and combine it with the high accuracy of cross-encoder. Bi-encoder model assigns a similarity score (between 0 and 1) to each chunk based on its relevance to the input query. We then sort the chunks by their scores and pick the top k (in our case, k=32) most relevant ones for further processing. In the second stage, we use a more accurate but computationally expensive cross-encoder model. We feed each of the top k chunks from the first stage, along with the original query, into the cross-encoder. Finally, we select the top m (in our case, m=3) most relevant chunks based on the cross-encoder scores and return them as context for our prompts. The detailed overview of the described process is depicted in Fig. 2.

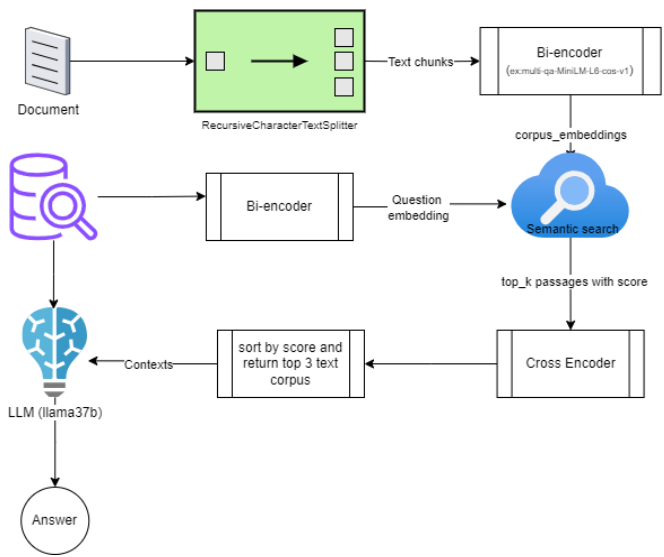

Fig. 2. Detailed insight into two-stage search process.

## IV. CASE STUDIES

### A. Automotive Compliance Chatbot

The goal of this case study is to provide a chatbot that would help automotive engineers during complex design phases, by giving them answers to compliance-related questions. In the first step, the user uploads a document which would be used as a reference for answers by LLM. In our proof-of-concept implementation, we were relying on the following documents related to automotive standards and reference requirements: ISO/TR 4804:2020[7], ISO/CD TS 5083[11], ISO 26262-1:2018[12] and AVC Consortium's reference specification[13]. While the first two are ISO standard norm documents, the third one is a set of good practices for automotive requirements related to various aspects of autonomous driving capabilities.

Fig. 2 shows our custom compliance chatbot based on AVC automotive requirements document [14] as reference.

### B. Automotive Design Copilot

The focus of the second case study is on leveraging the feedback given by LLM related to user-defined design in the form of copilot-alike tool. In this case study, we build upon our previous works [2] and[13]. The proposed workflow involving feedback to user's design decision during automotive software development from [13] is integrated with the code generation tools described in [2]. The illustration is given in Fig. 4.

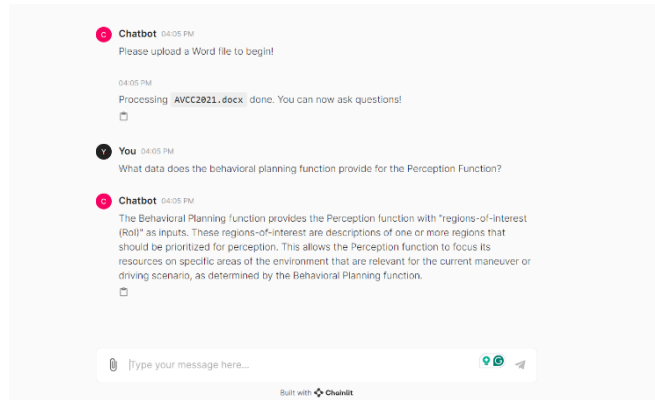

Fig. 3. Custom automotive design compliance chatbot.

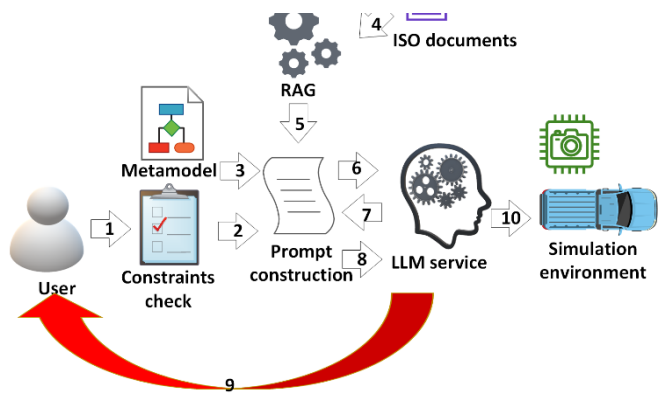

Fig. 4. LLM-based design feedback in automotive leveraging RAG: 1-user-specified model with respect to metamodel; 2-Failing OCL rules; 3-Metamodel specification as input 4-Load document into RAG 5-Document chunks with relevant context information; 6-Prompt 1 execution 7-LLM-generated suggestions 8-Prompt 2 execution; 9-User corrections of the model based on suggestions; 10-Code generation.

User designs vehicular system in model-driven manner, while interaction with RAG-empowered service gives hints and corrections - either on-demand (user asks what the resolution of camera for emergency should be brake) or automatically (checking camera resolution from model - parsing using Pyecore) and comparing the result from RAG. In case that correction is needed. In the first verification step, the user's design process outcome is Ecore model instance which is verified relying on OCL rules, as given in [2]. For that purpose, OCL rules had been previously generated using ISO standard document as input. In this step, the rules generated encapsulate general design guidelines, which are not trade secrets. After the model instance verification, as outcome we get the list of rules and their status - pass or fail. After that, the outcomes of verification are taken as input in order to construct prompts to RAG-enabled module which is able to insert the information coming from additional documents as part of querying context, such as vendor-specific good practices and parameter values which are aligned with existing ISO standards, but which are not publicly available or part of trade secrets. This way, we enable integration of relevant manufacturer-specific guidelines into the LLM-driven design workflow without exposing it.

The structure of the underlying prompt is the following. For each of the rules which are failing, we will have a prompt to LLM in this form:

*Prompt1: What to do to improve [model instance] if rule is not satisfied [failing rule] based on [retrieved context]?* (1)

The elements in square brackets are inserted within the query. Three inputs are required to be added to the query: 1) *model instance* - the whole instance model which was created by user and contains some rule violation 2) *failing rule* - textual representation of input used for OCL rule generation previously by LLM 3) *retrieved context* - output of RAG which contains the relevant information regarding the aspects of provided design which are not satisfied within the provided instance.

On the other side, once the model passes all the OCL constraints, another prompt is used in order to check whether the provided model instance provides complete set of requirements for the given scenario with respect to the selected reference document. In this case, the prompt is structured as follows:

*Prompt2: Are [scenario] requirements complete for [model instance] based on [retrieved context] as reference?* (2)

This way, we achieve two-step feedback workflow based on RAG and LLM to the user – first about constraint satisfaction and later about requirements completeness as well. Fig. 3 illustrates the proposed workflow and steps. Once user corrects the model instance, so it can pass both OCL rule and requirements completeness checks, the experiment code aiming CARLA[14] simulation environment will be generated, as described in [2].

## V. EXPERIMENTS AND EVALUATION

In this paper, we compared four Large Language Models (LLMs) with the aim to evaluate two aspects – accuracy (percentage of correct responses) and execution time while responding to a set of questions. The models involved into evaluation were: LLAMA3[15], [16], GPT-4o[17], and Mistral[18], all of which are the latest versions available as of June 2024.

Execution time and correction percentage are measured as average of 5 different prompts inspired by CeCaS (Central Car Server) project's workflow[19], based on ISO 26262-1:2018[12] document as input. Figure 5 shows the questions, and their correct answers extracted from ISO 26262-1:2018.

Regarding the question answer scores, the following scoring rules were applied for accuracy evaluation:

1 - the answer is correct, and the reasoning is also correct.

0.5 - the answer is correct, but the reasoning is incorrect.

0 - both the answer and reasoning are incorrect.

These experiments were conducted on a Google Collab Pro system[20] with the following configuration: runtime - Python 3; hardware accelerator - NVIDIA L4 GPU; system RAM - 53.0 GB; GPU RAM - 22.5 GB; disk space - 201.2 GB. We utilized the Langchain library to prepare and issue prompts to the OpenAI API. This library allowed us to efficiently handle the interaction with the API, streamline the process of generating responses, and manage the prompt engineering aspects of the experiment.

For comparing the performance of open-source Large Language Models (LLMs), we employed Ollama[21]. This tool enabled us to run these models locally, providing a means to directly measure and compare their runtime performance against the proprietary models accessed via the OpenAI API.

| Question | Answer |
|---|---|
| What is calibration data in the context of ISO26262? | Data that will be applied as software parameter values after the software build in the development process. |
| According to ISO 26262, can we consider the case of over voltage failure (3.50) due to lots of parts not meeting their specification for over voltage, as a common cause failure? | Yes, the example of over voltage failure is an example of common mode failure which is a case of common cause failure. |
| Can we consider a Kubernetes deployment yaml file as a configuration data? | Yes, since configuration data is only used to select code variants and it does not include the code itself, we can consider a Kubernetes deployment yaml file as configuration data. |
| How to specify whether a given failure is a cascading failure or a common cause failure? | Whether a given failure is a cascading failure or a common cause failure depends on the hierarchical structure of the elements and on the temporal behavior of the elements. Cascading failures propagate through the system, while common cause failures affect multiple elements due to a single root cause. |
| Suppose that a fault happened at time 3 and it got detected at time 6 and the diagnostic test time interval is 2. What is the FDTI (Fault Detection Time Interval)? | The FDTI is 3. FDTI is the time-span from the occurrence of a fault to its detection, independent of the diagnostic test time interval. |

Fig. 5. Questions and correct answers used for the experiment.

In Table II, the summary of accuracy-related evaluation is given. Based on the achieved results, we can notice the following for each of the models: 1) Llama3: achieved a total score of 2, provided correct answers with correct reasoning for one question and partial answers for two others; 2) Mixtral: scored 1.5, indicating challenges in providing accurate answers and reasoning, while fully answered only one question correctly 3) GPT-4.0: excelled with a total score of 4.5, consistently providing accurate answers and correct reasoning, making it the most reliable model in this comparison; 4) Mistral: matched Llama3 with a total score of 2, showing some capability but less consistency compared to GPT-4.0.

TABLE II QUESTION ANSWERING ACCURACY AND EXECUTION TIME EVALUATION

| Question | LLAMA3 | Mixtral | GPT-4o | Mistral |
|---|---|---|---|---|
| 1 | 0 | 0 | 1 | 0 |
| 2 | 0.5 | 0 | 1 | 0.5 |
| 3 | 0.5 | 1 | 1 | 0.5 |
| 4 | 0 | 0.5 | 0.5 | 0 |
| 5 | 1 | 0 | 1 | 1 |
| Overall score | 2 | 1.5 | 4.5 | 2 |
| Execution time [s] | 19.7 | 132.38 | 19.06 | 10.93 |

Overall, GPT-4.0 demonstrated superior performance in terms of both accuracy and reasoning, making it the most effective LLM among those evaluated in this study. However,

LLAMA3 and Mistral also show acceptable results, considering the ability to run them locally on lower capability hardware, which is one of the crucial advantages when it comes to adoption of LLMs in automotive domain.

Additionally, we also evaluate the average execution time per prompt for given questions in case of 3 runs, also shown in Table II (the last row). As can be seen, Mixtral is much slower than the other solutions, while Mistral exhibits the fastest response time.

We conducted additional simulations to assess the scalability of the proposed Retrieve and Re-Rank pipeline. Specifically, we varied the size of input documents to observe how it affects the performance of indexing and search processes. Table III summarizes the results, where we increased the number of pages to be indexed through the pipeline and measured two key metrics: *Indexing Time* which includes running the pipeline, and *Search Time*, which reflects the time required to search for relevant sections using the indexed chunks. As expected, increasing the size of the input documents leads to a proportional increase in indexing time, as this process involves computationally intensive operations such as chunking and model inference. However, the search time remains nearly constant, even as the document size grows. This is because the Bi-encoder and Cross-encoder models used in the pipeline are optimized for fast retrieval, regardless of document size. These findings highlight the scalability of the search process and emphasize the advantage of using RAG methods to maintain fast response times, even in large-scale Question-Answering systems.

TABLE III SCALABILITY OF RETRIEVE AND RE-RANK PIPELINE

| number of pages | Indexing time | Search time |
|---|---|---|
| 901 | 0.95 | 0.49 |
| 3031 | 5.21 | 0.51 |
| 5162 | 9.47 | 0.52 |
| 7293 | 13.62 | 0.52 |

## VI. CONCLUSION AND FUTURE WORKS

In this study, we explored the integration of Large Language Models (LLMs) with Retrieval-Augmented Generation (RAG) to enhance design and software development in the automotive industry. Our research demonstrates that LLMs, when augmented with RAG, can effectively address the challenge of handling non-disclosable data in automotive domains. We implemented two case studies: a standardization compliance chatbot and a design copilot, leveraging RAG to provide context-aware and accurate responses.

This study demonstrates that current locally deployable open-source Large Language Models (LLMs) are still behind the commercial ChatGPT's underlying GPT-4 when it comes to providing the necessary accuracy for question-and-answer tasks in the automotive industry using Retrieval Augmented Generation (RAG) technology. While the outcomes of models like LLAMA3 and Mistral seem more promising, they do not yet meet the stringent requirements for reliable performance in this specialized field. However, GPT-4 stands out by providing almost completely correct answers, indicating that proprietary models are currently more capable than open-source alternatives.

Future work should focus on improving the accuracy and reliability of open-source models to better support automotive applications by performing model fine tuning. Additionally, we will aim to extend the evaluation by testing additional open-source and commercial LLMs in order to identify the most effective models for various automotive applications. The effect of different bi-encoder and cross-encoder models on the performance of the RAG pipeline will also be evaluated, exploring models with varying architectures and training datasets to optimize retrieval and generation processes. Finally, the current manual assessment of LLM outputs against ground truth answers will be automated using semantic comparison models, such as cross-encoder models, to enhance accuracy and efficiency.

ACKNOWLEDGMENT

This research was funded by the Federal Ministry of Education and Research of Germany (BMBF) as part of the CeCaS project, FKZ: 16ME0800K.

REFERENCES

[1] N. Petrović and I. Al-Azzoni, "Model-Driven Smart Contract Generation Leveraging ChatGPT," in Advances in Systems Engineering, H. Selvaraj, G. Chmaj, and D. Zydek, Eds., Cham: Springer Nature Switzerland, 2023, pp. 387–396. doi: 10.1007/978-3-031-40579-2_37.

[2] N. Petrovic et al., "Synergy of Large Language Model and Model Driven Engineering for Automated Development of Centralized Vehicular Systems," Apr. 08, 2024, arXiv: arXiv:2404.05508. doi: 10.48550/arXiv.2404.05508.

[3] L. S. M. Netz, J. Michael, and B. Rumpe, From Natural Language to Web Applications: Using Large Language Models for Model-Driven Software Engineering. Gesellschaft für Informatik e.V, 2024. doi: 10.18420/modellierung2024_018.

[4] J. Dwight, Retrieval Augmented Generation. Independently published, 2024.

[5] Y. Gao et al., "Retrieval-Augmented Generation for Large Language Models: A Survey," Mar. 27, 2024, arXiv: arXiv:2312.10997. doi: 10.48550/arXiv.2312.10997.

[6] "Introduction to LangChain." Accessed: Sep. 16, 2024. [Online]. Available: https://python.langchain.com/v0.2/docs/introduction/

[7] ISO, ISO/TR 4804:2020 Road vehicles — Safety and cybersecurity for automated driving systems — Design, verification and validation, Geneva, Switzerland., 2020. Accessed: Sep. 16, 2024. [Online]. Available: https://www.iso.org/standard/80363.html

[8] "ChatGPT | OpenAI." Accessed: Sep. 13, 2024. [Online]. Available: https://openai.com/chatgpt/

[9] "sentence-transformers/multi-qa-MiniLM-L6-cos-v1 · Hugging Face." Accessed: Sep. 16, 2024. [Online]. Available: https://huggingface.co/sentence-transformers/multi-qa-MiniLM-L6-cos-v1

[10] "cross-encoder/ms-marco-MiniLM-L-6-v2 · Hugging Face." Accessed: Sep. 16, 2024. [Online]. Available: https://huggingface.co/cross-encoder/ms-marco-MiniLM-L-6-v2

[11] ISO, ISO/DTS 5083: Road vehicles — Safety for automated driving systems — Design, verification and validation, Geneva, Switzerland., 2024. Accessed: Sep. 16, 2024. [Online]. Available: https://www.iso.org/standard/81920.html

[12] ISO, ISO 26262-1:2018 : Road vehicles — Functional safety, Geneva, Switzerland., 2018. Accessed: Sep. 16, 2024. [Online]. Available: https://www.iso.org/standard/68383.html

[13] Autonomous Vehicle Computing Consortium, "Technical Report 001 Conceptual Architecture for Automated and Assisted Driving

Systems," Technical Report 001, Apr. 2021. [Online]. Available: https://avcc.org/tr001/

[14] "CARLA Simulator." Accessed: Sep. 16, 2024. [Online]. Available: https://carla.readthedocs.io/en/latest/

[15] "Introducing Meta Llama 3: The most capable openly available LLM to date," Meta AI. Accessed: Sep. 16, 2024. [Online]. Available: https://ai.meta.com/blog/meta-llama-3/

[16] "mixtral." Accessed: Sep. 16, 2024. [Online]. Available: https://ollama.com/library/mixtral

[17] "Hello GPT-4o." Accessed: Sep. 16, 2024. [Online]. Available: https://openai.com/index/hello-gpt-4o/

[18] "Bienvenue to Mistral AI Documentation | Mistral AI Large Language Models." Accessed: Sep. 16, 2024. [Online]. Available: https://docs.mistral.ai/

[19] K. Lebioda, V. Vorobev, N. Petrovic, F. Pan, V. Zolfaghari, and A. Knoll, "Towards Single-System Illusion in Software-Defined Vehicles -- Automated, AI-Powered Workflow," Mar. 21, 2024, arXiv: arXiv:2403.14460. doi: 10.48550/arXiv.2403.14460.

[20] "Google Colab." Accessed: Sep. 16, 2024. [Online]. Available: https://colab.research.google.com/

[21] "Ollama." Accessed: Sep. 16, 2024. [Online]. Available: https://ollama.com